\documentclass[12pt, draftclsnofoot, onecolumn]{IEEEtran}
\usepackage{amsmath, graphics,amssymb,epsfig,subfigure}
\usepackage{etoolbox}
\usepackage{cite,xcolor}
\usepackage{url}
\usepackage{algorithm}
\usepackage{color,soul}
\usepackage{cases}
\usepackage{graphicx,amsmath}
\usepackage{stfloats}
\usepackage{multirow}
\usepackage{comment}

\begin{document}
\title{Large Language Models for Knowledge-Free Network Management: Feasibility Study and Opportunities}
\author{Hoon Lee~\IEEEmembership{Member,~IEEE}, Mintae Kim, Seunghwan Baek, Namyoon Lee~\IEEEmembership{Senior Member,~IEEE}, Merouane Debbah~\IEEEmembership{Fellow,~IEEE}, and Inkyu Lee~\IEEEmembership{Fellow,~IEEE}
\thanks{

H. Lee is with the Department of Electrical Engineering and the Artificial Intelligence Graduate School, Ulsan National Institute of Science and Technology (UNIST), Ulsan, 44919, Korea.

M. Kim, S. Baek, N. Lee, and I. Lee are with the School of Electrical Engineering, Korea University, Seoul 02841, Korea (e-mail: inkyu@korea.ac.kr).

M. Debbah is with Khalifa University of Science and Technology, PO Box 127788, Abu Dhabi, UAE. \textit{(Corresponding authors: Inkyu Lee.)}
}}

\maketitle
\begin{abstract}
Traditional network management algorithms have relied on prior knowledge of system models and networking scenarios. In practice, a universal optimization framework is desirable where a sole optimization module can be readily applied to arbitrary network management tasks without any knowledge of the system.
To this end, knowledge-free optimization techniques are necessary whose operations are independent of scenario-specific information including objective functions, system parameters, and network setups. The major challenge of this paradigm-shifting approach is the requirement of a hyper-intelligent black-box optimizer that can establish efficient decision-making policies using its internal reasoning capabilities. This article presents a novel knowledge-free network management paradigm with the power of foundation models called large language models (LLMs). Trained on vast amounts of datasets, LLMs can understand important contexts from input prompts containing minimal system information, thereby offering remarkable inference performance even for entirely new tasks. 
Pretrained LLMs can be potentially leveraged as foundation models for versatile network optimization. 
By eliminating the dependency on prior knowledge, LLMs can be seamlessly applied for various network management tasks. 
The viability of this approach is demonstrated for resource management problems using GPT-3.5-Turbo. Numerical results validate that knowledge-free LLM optimizers are able to achieve comparable performance to existing knowledge-based optimization algorithms.
\end{abstract}

\section{Introduction}
Requirements for next-generation wireless communication systems become significantly complicated and detailed, necessitating comprehensive optimization across various aspects. This invokes in-depth knowledge in developing accurate models about individual networking applications including hardware impairments, propagation environments, and network architectures. 
With predefined models, traditional model-based optimization strategies have readily been employed for desired environments, but they are tailored only to specific systems.

Triggered by rapid developments in artificial intelligence (AI) technologies \cite{HLee:19JSAC,DDPG}, the data-driven optimization method using neural networks can potentially eliminate model-dependent computations as it relies solely on training datasets. Nevertheless, this data-driven optimization approach can result in data overfitting during the training process, leading to generalization errors on unseen data. 
Consequently, both the model-based and AI-driven solutions lack the adaptability for other network applications with new features and requirements.

A versatile network management strategy is imperative to adapt to dynamically evolving network services without additional optimization and tuning processes \cite{Wang:23}. 
A primary requirement is to remove the need for system-specific knowledge, such as channel state information (CSI), mathematical descriptions, and network environments. In such knowledge-free optimization paradigms, a single optimization module can be readily applied to various network management problems without further modifications. To achieve this goal, it is essential to develop a universal optimizer equipped with sophisticated decision-making capabilities, relying solely on its internal intelligence. As a result, outstanding reasoning processes can identify desired characteristics of network management tasks from very limited prior knowledge. 

Large language models (LLMs) based on generative pretrained transformer (GPT) can be exploited as foundation components for such applications. Trained on extensive datasets, LLMs excel at extracting meaningful semantics from input prompts using their own internal logic, which is a crucial feature for universal network management.
The viability of the LLM as an optimizer has been demonstrated in various cases \cite{OPRO,GD,LMEA,LEO,MOEAD,MALLM,LLMTelecom}. 
These methods, however, suffer from fundamental challenges of generative AI models incurring performance degradations \cite{OPRO} or unintended behaviors \cite{LEO,MALLM}, diminishing their practical utility. To resolve these difficulties, several approaches incorporate LLMs as components of existing model-based algorithms \cite{LEO,LMEA,GD}, making them tailored only to specific problems. Thus, the full potential of LLMs as standalone and knowledge-free optimizers remains unaddressed.

In this article, we introduce a new knowledge-free universal optimization framework for wireless resource management tasks by leveraging the power of LLM. The key feature of the proposed framework is its ability to optimize without requiring any mathematical models and additional training. We harness multiple LLMs to build collective intelligence that cooperatively tackles resource allocation problems.
The effectiveness of the proposed framework is exemplified by power control problems in interference networks. Numerical results confirm that the proposed method outperforms conventional LLM optimizer techniques and exhibits performance comparable with existing algorithms. We also present several open challenges and future research opportunities associated with the proposed universal optimization method.

This article first discusses key features required for knowledge-free optimizers and promotes LLMs as foundation components for such innovative frameworks. An overview of state-of-the-art LLM optimizer techniques is presented along with their fundamental limitations. A novel multi-LLM collaboration approach is proposed to address resource management tasks. The effectiveness of the proposed framework is demonstrated via numerical results using GPT-3.5-Turbo. The article is terminated with concluding remarks and research directions.

\section{Knowledge-Free Network Management}
As next-generation wireless communication systems evolve, their applications become increasingly diversified, aiming to incorporate heterogeneous services including wireless sensing, positioning, and AI computations. 
This solicits versatile and self-organizing network management algorithms capable of automatically handling a vast array of configurations and services without problem-specific knowledge \cite{Wang:23}. This section presents the concept of knowledge-free optimization, which is followed by the potential of using LLMs as network optimizers.

\subsection{Requirements for knowledge-free optimizer}
Traditional optimization techniques such as model-based algorithms and AI-driven solutions 
are prone to overfitting to specific scenarios.
Such a rigid property is induced by the dependency on prior knowledge about systems. In contrast, a knowledge-free network optimization method does not require any knowledge of the network parameters or specific network conditions. As a result, this method has a high potential to achieve autonomy and universality for arbitrary network optimization problems. Notwithstanding its promise, several critical challenges must be addressed to realize this paradigm-shifting optimization framework, which are summarized as follows:
\begin{itemize}
    \item \textit{Model-free:} System models including formulas of objective/constraint functions, network parameters, channel models, and input prior have been widely exploited as requisite for deriving advanced network management solutions. This makes existing algorithms and AI models overfit to specific tasks and systems, lacking flexibility and making generalization not feasible on unseen scenarios. These model and data dependent characteristics should be eliminated in designing universal network optimization policies.
    \item \textit{Hyperparameter-free:} Model-free algorithms such as meta-heuristic and deep reinforcement learning (DRL) methods \cite{DDPG} typically entail a number of hyperparameters, e.g., the architecture of AI models, input features, and training loss/reward functions. Careful crafting of such components is required to solve network problems, but this results in solutions dedicated to only specific scenarios. Eliminating the knowledge about hyperparameters is essential for versatile network optimization.
    \item \textit{Tuning-free:} It is highly difficult to apply AI models trained over specific systems to other configurations with different characteristics. Although recent AI models have evolved to enhance their adaptability to unseen datasets, they often need retraining or fine-tuning processes with new datasets. The dependency on such additional tuning phases along with the knowledge biased to training datasets should be removed so that the resulting optimizer can exhibit high-level generalization ability.
\end{itemize}

LLMs are perfect candidates for resolving model-free, hyperparameter-free, and tuning-free challenges, acting as a universal network optimizer. The power of LLMs in language understanding, planning, and reasoning is potentially advantageous for appropriate decision-making components. Trained with a vast amount of datasets, LLMs possess their own built-in knowledge base. This leads to an in-context learning (ICL) feature capable of inferring core characteristics of unseen problems from several examples provided in input prompts. Thanks to such properties, LLMs can be readily applied to new tasks. These unique features are beneficial for creating versatile network management policies without additional fine-tuning processes.

\subsection{Large language model as knowledge-free optimizer}

\begin{figure*}
    \centering
    \includegraphics[width=\linewidth]{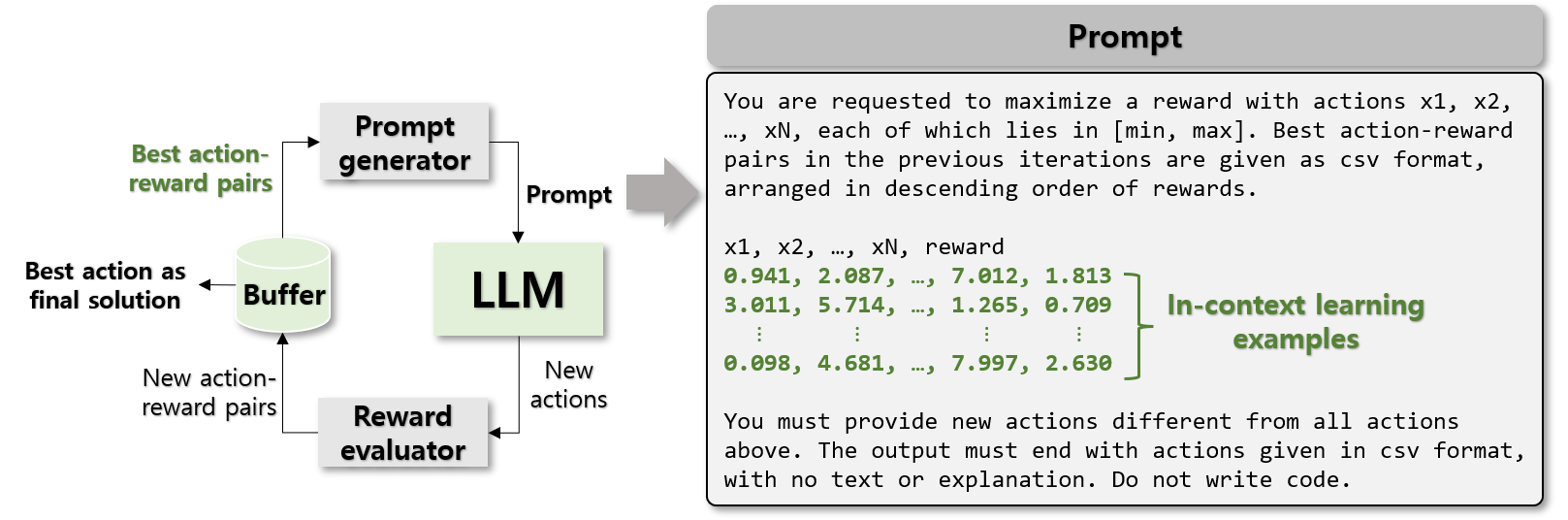}
    \caption{Knowledge-free LLM optimizer architecture.}
    \label{fig:fig1}
\end{figure*}

Fig. \ref{fig:fig1} illustrates an architecture of knowledge-free LLM optimizer presented in \cite{OPRO} where an LLM is utilized as a black-box operator to generate solutions through prompting. 
An input prompt includes descriptions of target optimization tasks, constraints, and desired operations requested for LLM in natural language form. This instructs the LLM to take effective actions with enhanced reward values and output them in the desired format. No knowledge-based details, such as mathematical models and computation guidelines, are included in the prompt, thereby ensuring a knowledge-free implementation.

To harness the autonomy of LLMs, the ICL approach is adopted which provides several examples in natural language form. This is particularly efficient for aligning LLMs to new tasks without the need for fine-tuning. An initial prompt provides a set of randomly selected actions and the corresponding reward values.
Then, these ICL examples are formed for LLMs to understand the inherent nature of optimization problems. By leveraging the ICL process, LLMs can progressively produce more effective actions that yield higher reward values. 
The sequence of actions taken by the LLMs, along with the associated rewards measured by a reward evaluator, are stored in a buffer.
In the subsequent step, the best actions in history are extracted from the buffer and are fed back to the LLM through the input prompt. This self-feedback enables the LLM to iteratively refine its actions, learning from past rewards to improve future decisions. Such a multi-step reasoning mechanism realizes the chain-of-thought mechanism \cite{CoT}, which prompts LLMs through a series of short sentences containing intermediate reasoning steps. This method facilitates the high-level inference ability of LLMs without any problem-specific information. 

Conventional LLM optimizer technique \cite{OPRO} has been examined in simple convex optimization tasks, and its viability for handling nonconvex network management tasks has not yet been studied adequately. 

\section{Surveys on LLM Optimizer Techniques}



\subsection{State-of-the-arts}

\begin{table*}[]
\centering
\caption{LLM-based optimization techniques ($\bigcirc$: positive, $\bigtriangleup$: neutral, $\times$: negative)}\label{tab:tab1}
\begin{tabular}{|c||c|c|c|c|c|}
\hline
Methods                           & Applications & Model-free & Hyperparameter-free & Tuning-free & LLM collaboration \\ \hline\hline
\cite{OPRO} & Linear regression, TSP      & $\bigcirc$                                                         & $\times$                                                              & $\bigcirc$                                                          &  $\times$                                                                \\ \hline
\cite{GD}        & Linear regression           & $\times$                                                       & $\times$                                                             & $\bigcirc$                                                          & $\times$                                                                   \\ \hline
\cite{LMEA}      & TSP          & $\bigcirc$                                                       & $\times$                                                             & $\bigcirc$                                                            & $\times$                                                                  \\ \hline
\cite{LEO}       & Multi-objective optimization          & $\times$                                                       & $\times$                                                          & $\bigcirc$                                                            & $\triangle$                                                                \\ \hline
\cite{MOEAD}     & Multi-objective optimization          & $\times$                                                         & $\times$                                                            & $\times$                                                         & $\times$                                                                \\ \hline
\cite{MALLM}     & Wireless resource allocation          & $\times$                                                       & $\bigcirc$                                                              & $\bigcirc$                                                         &  $\triangle$                                                               \\ \hline
Proposed                          & Wireless resource allocation          & $\bigcirc$                                                         & $\bigcirc$                                                              & $\bigcirc$                                                         & $\bigcirc$                                                                 \\ \hline
\end{tabular}
\end{table*}

Table \ref{tab:tab1} compares several existing LLM optimizer techniques.
The LLM optimizer was first introduced in \cite{OPRO} for solving linear regression and traveling salesman problems (TSP). Similar to meta-heuristic algorithms and DRL methods, the exploration-exploitation dilemma plays a critical role in constructing proper LLM optimizers. To control such a tradeoff, the temperature hyperparameter is adjusted manually, which determines the creativity of LLMs. A high temperature improves the creativity and diversity of LLMs, which helps to explore new solution spaces. 
On the contrary, LLMs with low temperature values become more focused on certain action candidates. This necessitates a careful hyperparameter optimization for individual optimization problems. 

Recent efforts have addressed the shortcomings of \cite{OPRO} by incorporating model-dependent operations \cite{GD,LMEA,LEO,MOEAD,MALLM}. They resort to problem-specific information for crafting input prompts \cite{GD,LMEA,LEO,MALLM} or generating ICL examples using traditional optimization algorithms \cite{LMEA,LEO}. Such operations successfully improve the exploration ability, thereby enhancing their decision-making performance without controlling the temperature parameter.
Mathematical formulations for reward functions are explicitly provided within input prompts \cite{GD}. LLMs are instructed to calculate gradients of reward functions and update actions in descent directions using the prompted step size. 
Although it is superior to \cite{OPRO}, these model-dependent input sentences need hyperparameter optimization through prompting. A similar approach has been adopted in \cite{LMEA} to tackle TSP. Inspired by meta-heuristic algorithms, this method utilizes model-based prompts that direct LLMs to engage in evolutionary operations.
The temperature parameter is updated over reasoning steps of LLMs, which invokes additional fine-tuning of such update periods. 

LLM optimizers were leveraged for solving multi-objective optimization problems \cite{LEO,MOEAD}. These methods adopt model-based computation procedures for generating prompts \cite{LEO} and ICL examples \cite{LEO,MOEAD}. LLMs act as surrogate operators of backbone meta-heuristic algorithms, and the remaining components still resort to model-dependent operations, requesting careful hyperparameter tuning. The scheme in \cite{MOEAD} employs LLM optimizers to produce training samples for other AI models. Therefore, the overall performance is highly dependent on the fine-tuning phase.

To enhance the exploration ability, random noise is injected into actions in \cite{LEO}, which needs additional hyperparameter optimization for noise variances. Besides, this method allows coordination between two LLMs. One serves as a normal LLM optimizer that identifies enhanced actions, whereas the other acts as an explorer tasked to find exploration samples away from current action candidates. These exploration samples are directly fed into the LLM optimizer, enabling it to expand its search into new areas. This coordination policy successfully improves the exploration ability, but it is confined to using only two LLMs.
Moreover, to effectively determine potential exploration samples, it requires carefully designed model-based prompts. Without these prompts, it might generate irrelevant or suboptimal actions.

Resource allocation tasks in multi-user networks have been recently investigated in \cite{MALLM}. A multi-agent setup was considered where individual LLM optimizers are responsible for the decisions of different users. These LLMs communicate by exchanging local decisions through input prompts. However, collaboration in this context is not intended for high performance gains. Detailed descriptions of target tasks, e.g., system models, CSI, and mathematical formulas of objective functions, are included in input prompts. Such a model-based prompting strategy guides LLM optimizers in accomplishing a particular problem with a certain system input. However, the viability of other setups cannot be demonstrated due to the limited adaptability of knowledge-based prompting.

Compared to these existing techniques, the proposed framework does not rely on model-dependent operations and additional training/tuning steps of AI models or hyperparameters. Exploiting the collective intelligence of LLMs is a key enabler for realizing the proposed knowledge-free LLM optimizers.

\subsection{Challenges for Knowledge-Free LLM Optimizers}


Several challenges associated with knowledge-free LLM optimizers are discussed as follows:

\subsubsection{Randomness}
Output responses of LLMs are random, meaning that they can vary even with identical inputs. This randomness becomes more pronounced when optimization problems involve the time-varying dynamics of wireless networks. 
Consequently, achieving a stable solution with the LLM optimizer becomes challenging, and 
this variability may lead to the fluctuation of the solutions.


The self-consistency technique \cite{CoT-SC} can overcome such an issue. This method employs multiple LLMs in parallel and ensembles their outputs to stabilize randomized behaviors. 
Among actions taken by several LLM optimizers, the best action achieving the maximum reward can be chosen as the final solution. However, this naive multi-LLM coordination strategy, which simply takes the best action among candidates, still suffers from performance degradation and requires a number of reasoning steps to attain reliable solutions.
    
\subsubsection{Mode collapse}
The mode collapse issue is a common problem of generative AI. Instead of generating creative responses, LLMs often produce certain outputs that only care about reducing training loss function instead of improving the quality of responses. In network management applications, LLM optimizers adopt a myopic strategy that focuses only on the proximity of previously explored successful actions. Thus, they often fail to identify optimal solutions that are potentially located outside current search spaces. Consequently, the performance quickly gets saturated and exhibits no further improvement.

The mode collapse limits the exploration ability of LLM optimizers. As evidenced from DRL algorithms \cite{DDPG}, a simple but powerful solution is to employ randomized policies such as noise injection \cite{LEO}. However, this reintroduces the need for hyperparameter optimization steps tailored to each given network scenario.
    
\subsubsection{Hallucination} 
LLMs frequently yield meaningless patterned numeric, e.g., repeating decimals and integers \cite{OPRO,LEO}. This phenomenon is referred to as the hallucination where LLMs produce unfaithful information while superficially maintaining the desired context~\cite{Hallucination}. To address this, the self-reflection mechanism \cite{Hallucination} can be employed for reducing the hallucination. This approach modifies incorrect sentences and feeds the refined prompts back to LLMs until the hallucinations are mitigated. However, this strategy relies on background datasets to detect and correct hallucination sentences, and thus it is not suitable for developing knowledge-free LLM optimizers.

The mode collapse and hallucination issues are closely related to the exploration ability of LLM optimizers. 
Repeatedly chosen suboptimal and hallucinatory actions make LLM optimizers stuck to undesirable solutions with poor performance. 
This issue becomes more severe when handling nonconvex network management tasks, which require high-level exploration to find globally optimum actions. 
There have been a number of efforts to improve the exploration capability of LLM optimizers. 
However, as presented in Table \ref{tab:tab1}, these conventional methods request knowledge-dependent features such as model-based operations \cite{GD,LMEA,LEO,MOEAD,MALLM}, hyperparameter optimization \cite{OPRO,LEO}, and additional training steps \cite{MOEAD}, which are unsuitable for knowledge-free network management.

\section{Proposed Collaborative Multi-LLM Optimizer}


We propose a knowledge-free LLM optimizer designed to solve resource management tasks in wireless networks. 
Inspired by \cite{CoT-SC}, we leverage multiple LLMs working together to mitigate their inherent randomness. This offers a multi-LLM diversity that allows us to investigate diverse actions produced by individual LLMs, potentially leading to improved exploration ability. A naive multi-LLM architecture in \cite{CoT-SC}, which only takes the best LLM with the highest reward, simply discards inferior LLMs that might generate better exploration samples.

To fully harness the collective intelligence of LLMs, it is essential to develop a proper LLM collaboration strategy. Our approach aggregates diverse responses of LLMs and exploits them as ICL examples. This policy allows the LLMs to dynamically and adaptively explore various solutions taken by other LLMs, progressively improving the effectiveness of actions and increasing reward values over time. As a result, the proposed approach can overcome the limited exploration ability, which is related to the mode collapse and hallucination issues, in a knowledge-free manner without additional fine-tuning processes.
The harmonization of collective LLMs and their dynamic coordination controls is an essential building block for realizing knowledge-free LLM optimizers.

\subsection{Collaborative multi-LLM optimizer architecture}

\begin{figure*}
    \centering
    \includegraphics[width=.9\linewidth]{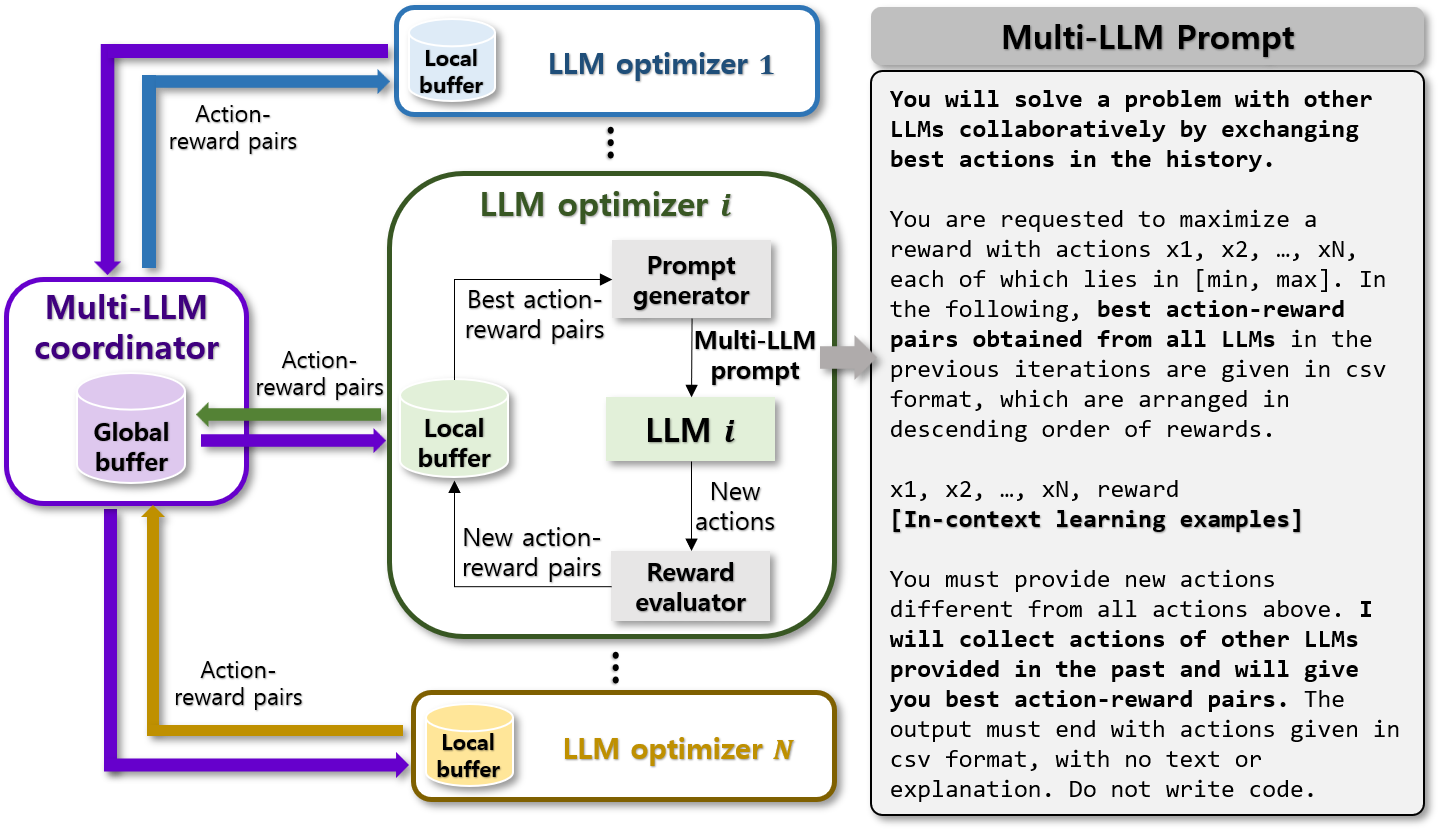}
    \caption{Proposed collaborative multi-LLM network optimizer.}
    \label{fig:fig3}
\end{figure*}

Fig. \ref{fig:fig3} illustrates the proposed LLM optimization architecture that collaboratively performs network optimization tasks through inter-LLM interaction. The proposed architecture comprises two major components: LLM optimizers and a multi-LLM coordinator.
Each LLM optimizer is designed for network management problems via prompting. Actions generated by a group of LLM optimizers are exchanged with each other through the multi-LLM coordinator. The details of these components are discussed below.

\subsubsection{LLM optimizer}
Unlike the conventional method \cite{LEO} which only uses two LLMs, the proposed framework exploits an arbitrary number of LLMs. Each individual LLM optimizer consists of a prompt generator, a reward evaluator, and a local buffer. The prompt generator yields the multi-LLM prompt containing desired instructions for LLMs. Compared to existing single LLM techniques \cite{OPRO}, the proposed multi-LLM prompt includes additional descriptions about inter-LLM collaboration protocols, which are highlighted by boldface letters in Fig.~\ref{fig:fig3}. The prompt begins with an explanation of optimization tasks followed by a brief introduction to the interaction policy. Its detailed operations are given along with ICL samples of the best-action reward pairs, informing LLMs that these actions correspond to the most effective candidates among all LLMs. 

The reward evaluator interacts with an environment, e.g., mobile users and networks, to assess the reward values of actions. The resulting action-reward pairs are stored in the local buffer, which organizes them in descending order of reward values. At each iteration, the best action-reward pairs are extracted from the local buffer and are fed to the prompt generator to provide ICL samples to the LLM. At the same time, these are shared with the multi-LLM coordinator to facilitate interaction with other LLM optimizers.

\subsubsection{Multi-LLM coordinator}
This module plays a pivotal role in coordinating LLMs. It is equipped with a global buffer, which harnesses the collective experiences of multiple LLM optimizers. This unit consolidates the most effective action-reward pairs cached in the local buffers. Such a comprehensive knowledge base implements enhanced decision-making policies at all LLMs.

LLMs interact with each other by exchanging their action-reward pairs. The multi-LLM coordinator governs the direction of such information flows between local and global buffers according to the performance of individual LLM optimizers.  
If a certain LLM optimizer takes improved actions compared to those stored in the global buffer, 
these 
are pushed from the local buffer to the global buffer. On the contrary, the best action-reward pairs are popped from the global to local buffers of inferior LLM optimizers that cannot find improved actions. Consequently, the multi-LLM coordinator establishes dynamic switching operations that bridge the local and global buffers.

The proposed multi-LLM collaboration protocol keeps collecting the most effective actions from superior LLM optimizers that improve rewards. Therefore, the global buffer contains up-to-date samples with the largest rewards. This can be viewed as the ensemble learning method which combines several AI models to yield the best inference results. In our case, we synthesize the collective intelligence of LLMs to form elite actions.
These are distributed to inferior LLM optimizers that fail to improve reward values. These updated samples, which possess higher rewards than those stored in local buffers, are utilized as ICL examples in the subsequent iteration. Consequently, action-reward pairs selected from the global buffer act as corrected sentences of the self-reflection mechanism \cite{Hallucination}, which dynamically adjusts input prompts to modify hallucinatory responses.

In addition, porting successful actions from the global buffer to local buffers can be viewed as the noise injection procedure \cite{LEO}. This is beneficial for inferior LLM optimizers which are stuck to suboptimal actions. We can help these LLMs stimulate exploration and prevent immature convergence. It is noted that the proposed approach generalizes the conventional method in \cite{LEO} where the global buffer always collects new actions of all LLM optimizers and dispatches the best actions to all local buffers in every iteration. Unlike such a passive coordination policy, the proposed multi-LLM coordinator dynamically adjusts the period of such action exchanges based on the states of individual LLM optimizers. This ensures adaptive exploration-exploitation controls, thus fully facilitating the autonomy of the overall framework without additional hyperparameter optimization.

\subsection{Performance Evaluation}

We evaluate the proposed framework for power control tasks that maximize energy efficiency (EE) and spectral efficiency (SE) of three-cell interference channels. The transmit power budget and circuit power consumption of base stations are set to 10 W and 1 W, respectively. The variance of the additive Gaussian noise is fixed to unity for the Rayleigh fading channel. All simulations are implemented with GPT-3.5-Turbo. LLM optimizers are first prompted by 5 random initial actions. They are instructed to determine 5 new transmit power actions based on 10 ICL examples. Unless otherwise stated, 5 LLMs are employed and their best-case performance is examined. The proposed approach is compared with the following baseline techniques.
\begin{itemize}
    \item \textit{Local optimal:} Performance upperbounds are generated using existing model-based algorithms. Local optimum solutions for EE and SE maximization problems are obtained from the fractional programming and the WMMSE method, respectively.
    We choose the best performance obtained with 5 random initial points.
    \item \textit{Passive collaboration (PC) \cite{LEO}:} This method corresponds to a special case of the proposed dynamic multi-LLM coordination strategy. All LLMs simply share their new actions in every iteration. Such a passive inter-LLM interaction rule, which has been proposed in \cite{LEO}, prompts the most effective actions to all LLMs simultaneously regardless of their individual capabilities. Consequently, it cannot control the exploitation-exploration behavior of LLM optimizers.
    \item \textit{No collaboration (NC) \cite{OPRO}:} No exchanges of action-reward pairs among LLM optimizers are allowed during iterations. 
    Once completed, the best action that achieves the maximum performance is selected among candidate solutions generated by all LLMs. Thus, this is a straightforward extension of the conventional single LLM optimizer \cite{OPRO} combined with the self-consistency technique \cite{CoT-SC}.
    \item \textit{Brute-force:} Random actions are uniformly selected at each iteration whose population is the same as those determined by LLM optimizers.
\end{itemize}


\begin{figure}
\centering
    \subfigure[EE]{
        \includegraphics[width=.7\linewidth]{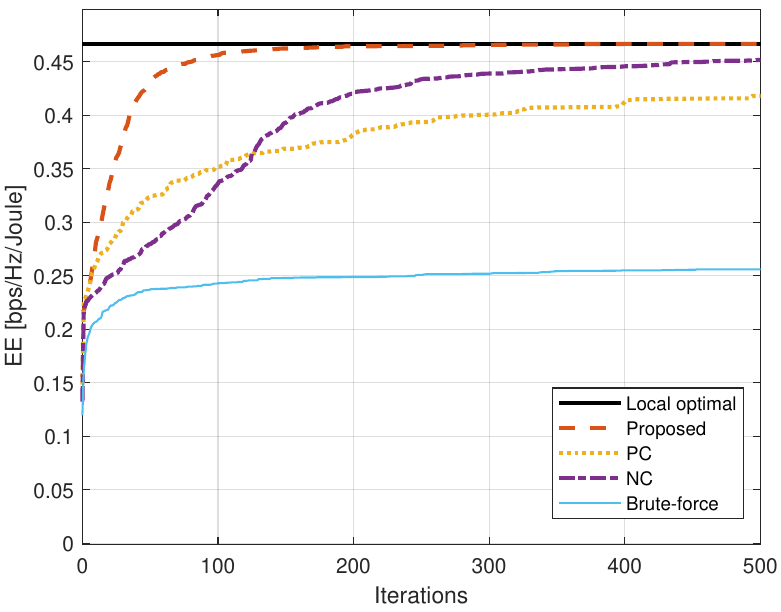}
    }
    \subfigure[SE]{
        \includegraphics[width=.7\linewidth]{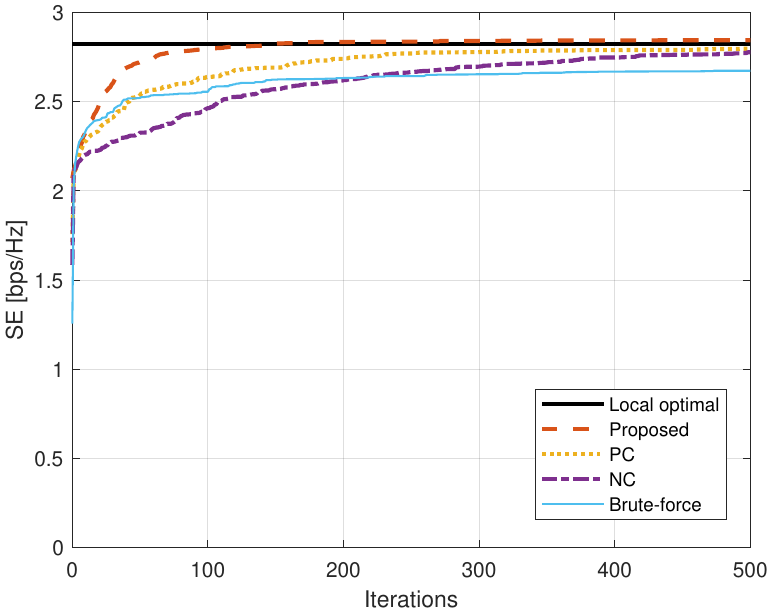}
    }
    \caption{Performance comparison of various LLM optimizer techniques.}
    \label{fig:fig4}
\end{figure}

Fig. \ref{fig:fig4} exhibits the EE (Fig. \ref{fig:fig4}(a)) and SE (Fig. \ref{fig:fig4}(b)) performance in terms of iterations. For all optimization metrics, the proposed method can achieve the performance upperbound with significantly fewer iterations than other LLM optimizer methods. 
For the EE maximization problem, the proposed method provides almost identical performance to the local optimum algorithm at the 100-th iteration, whereas both the PC and NC schemes fail to approach the upperbound within 500 iteration steps. This confirms the effectiveness of the proposed dynamic LLM collaboration. In addition, the PC is even worse than the NC after the intermediate iteration, proving that a passive LLM coordination \cite{LEO} possibly brings severe performance degradations. 
Note that the PC simply dispatches the most effective actions to all LLMs, and thus it forces all LLM optimizers to explore the identical action space regardless of their individual capabilities. Consequently, the exploration-exploitation tradeoff of each LLM optimizer cannot be controlled properly. In the beginning, the PC enhances the exploration ability of all LLMs, thereby accelerating the convergence. However, the exploration would be harmful after intermediate iterations where the exploitation becomes more important. For this reason, the performance of the PC method is saturated to a lower EE value. In contrast, the proposed framework dynamically adjusts the exploration level of individual LLM optimizers according to their current states. As a result, the proposed mechanism results in fast convergence and improved performance in all simulated cases.

In the SE case, the performance of the NC method improves slowly compared to the brute-force search. This infers that LLM optimizers require a number of ICL examples to determine their own decision-making policies. Through a valid multi-LLM collaboration, such a cold-start behavior can be resolved by allowing LLMs to observe actions explored by others. As a result, both the proposed scheme and the PC method successfully accelerate the convergence speed compared to the NC method. 

\begin{figure}
    \centering
    \includegraphics[width=.7\linewidth]{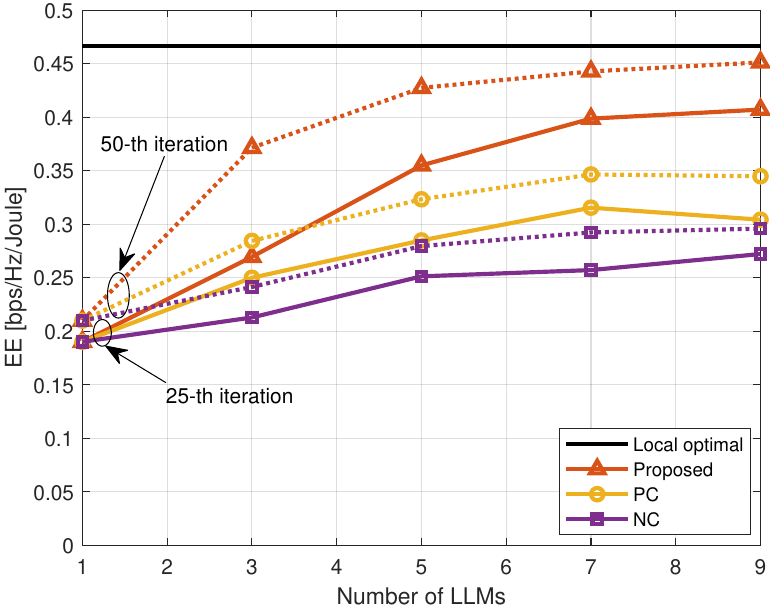}
    \caption{EE performance with respect to the number of LLMs.}
    \label{fig:fig5}
\end{figure}

The impact of the number of LLMs is examined in Fig.~\ref{fig:fig5} which compares the EE performance. The performance of the proposed method gradually increases with the number of LLMs. On the contrary, for the PC and NC, the multi-LLM gain becomes diminished, especially at the 25-th iteration. For the proposed framework, 5 LLMs are sufficient to reach $90\ \%$ of the local optimum performance at the 50-th iteration, whereas the PC requires more than 9 LLMs to achieve similar performance. This validates the advantages of the proposed collaboration strategy in leveraging multiple LLMs effectively.


\section{Future Research Directions}

\subsection{Theoretical analysis of LLM optimizer}
To exploit the full potential of LLM optimizers, it is essential to investigate their theoretical performance such as convergence and optimality. 
LLM optimizers can be interpreted as sequential decision makers that determine candidate actions with given historical samples. Therefore, rigorous approaches in reinforcement learning and evolutionary algorithms, along with those in language models, can be extended to derive theoretical guidelines for LLM optimizers.

\subsection{Constrained LLM optimizer}
Model-free prompting might not fulfill complicated network constraints such as quality-of-service requirements that are normally coupled with CSI. This issue can be resolved by the primal-dual learning strategy \cite{HLee:19JSAC}. Two distinct LLM optimizers can participate in the minimax optimization of Lagrangian by generating primal and dual solutions. With such an adversarial inter-LLM collaboration, a knowledge-free constrained optimizer can be established. 

\subsection{Warm-start LLM} 
To further accelerate the convergence, warm-starting techniques can be employed \cite{WS}. Initial ICL samples are generated so that LLM optimizers can identify global optimum points quickly. A distinct LLM can be prompted to produce warm-start samples that can speed up LLM optimizers. Developing a proper interaction strategy with such an LLM initializer would be a primary challenge.

\subsection{On-device LLM}
Due to high computation power for executing inference, GPT-enabled LLM optimizers highly resort to cloud implementation. To realize them at practical wireless devices, e.g., internet-of-things nodes, it is essential to leverage on-device LLMs with much fewer parameters. However, these shallow models often degrade inference performance. To address this, the mixture-of-expert concept can be employed to leverage a group of lightweight LLMs \cite{MoE}, shown to outperform GPT with reduced complexity.

\subsection{Decentralized LLM optimizer}
The proposed framework requires a centralized structure. In contrast, it is desirable for knowledge-free LLM optimizers to be distributed over cloud, edge, and devices. To facilitate these decentralized LLMs, their interaction protocols should be investigated. 
LLMs would be allowed to exchange text messages describing their optimization policies. By doing so, fully self-organizing network management is viable.

\subsection{Heterogeneous LLM collaboration}
Future networks may consist of heterogeneous LLMs with different computing abilities, distinct model structures developed by multiple companies, and diverse corpus in pertaining phases. These LLMs can be co-located at a certain node or distributed over networks. The development of self-organizing management strategies with heterogeneous LLM optimizers would be an ultimate challenge.

\section{Concluding Remarks}
This article has proposed a new knowledge-free optimization framework by leveraging the collective intelligence of LLMs. A primary goal of this approach is to develop a universal network optimizer whose operations are independent of problem-specific information. To this end, GPT-enabled LLM optimizers have been exploited to identify efficient resource allocation solutions. A novel multi-LLM collaboration strategy successfully handles inherent challenges of LLMs. The proposed mechanism dynamically controls the exploitation-exploration ability of individual LLMs, thereby enhancing network management performance. 
Numerical results have demonstrated the effectiveness of the proposed dynamic LLM coordination policy over conventional schemes. Notwithstanding its potential, several critical issues remain in utilizing the knowledge-free LLM optimizer. These include incorporating the capability to handle versatile optimization constraints, enhancing convergence speed, reducing complexity for on-device operations, and facilitating heterogeneous LLM collaboration.

\bibliography{arXiv}
\bibliographystyle{ieeetr}

\end{document}